\documentclass[showpacs,pre]{revtex4}
\usepackage{graphicx}

\usepackage[usenames]{color}
\usepackage{cancel}
\usepackage{hyperref}
\usepackage{bbm}
\usepackage[normalem]{ulem}
\usepackage{amsfonts}
\usepackage{amsmath}
\usepackage{graphicx}
\usepackage{epstopdf}
\usepackage{url}
\newcommand{\rcr}{r^*}
\newcommand{\bra}[1]{\left(#1\right)}
\newcommand{\brb}[1]{\left[#1\right]}

\newcommand{\bre}[1]{\left\{#1\right\}}

\begin{document}

\title{Solution space structure of random constraint satisfaction problems with growing domains }

\author{Wei Xu$^1$}
\author{Pan Zhang$^{2,3}$}
\author{Tian Liu$^4$}
\email{lt@pku.edu.cn}
\author{Fuzhou Gong$^5$}
\affiliation{
$^{1}$School of Mathematics and Systems Science, Beihang University, Beijing 100191, China\\
$^{2}$Santa Fe Institute, Santa Fe, New Mexico 87501, USA\\
$^{3}$State Key Laboratory of Theoretical Physics, Institute of Theoretical Physics, Chinese Academy of Sciences, Beijing 100190, China\\
$^{4}$Key Laboratory of High Confidence Software Technologies, Ministry of Education, School of EECS, Peking University, Beijing 100871, China\\
$^{5}$Institute of Applied Mathematics, Academy of Mathematics and System Science, Chinese Academy of Sciences, Beijing 100190, China
}

\date{\today}

\begin{abstract}


In this paper we study the solution space structure of model RB, a standard prototype
of Constraint Satisfaction Problem (CSPs) with growing domains. Using rigorous
the \textit{first} and the \textit{second moment method}, we show that in the
solvable phase close to the satisfiability transition, solutions are
clustered into exponential number of well-separated clusters, with each cluster contains sub-exponential number
of solutions. As a consequence, the system has a clustering (dynamical) transition but
no condensation transition.
This picture of phase diagram is different from
other classic random CSPs with fixed domain size, such as random $K$-Satisfiability (K-SAT) and
graph coloring problems, where condensation transition exists and is distinct from
satisfiability transition.
Our result verifies some non-rigorous results obtained using cavity method from spin glass theory.
\end{abstract}

\pacs{89.75.Fb, 02.50.-r, 64.70.P-, 89.20.Ff}

\maketitle

\section{Introduction}
Constraint satisfaction problems are defined as a set of discrete variables whose
assignments must satisfy a collection of constraints.
A CSP instance is said to be satisfiable if there exists a solution, i.e. an assignment to all variables that satisfies
all the constraints. The core question to CSPs is to decide whether a given instance is satisfiable.
CSPs have been studied extensively in mathematics and computer
science, and play an important role in the computational complexity theory.
Most of the interesting CSPs, such as boolean K-satisfiability problems and graph coloring problems,
are belong to class of NP-complete: in the worst case the time required to decide whether there
exists a solution increases very quickly as the size of the CSP grows.

In recent years, there are many interests on the average case complexity of CSPs, which
study the computational complexity of random ensembles of CSPs. It also has drawn considerable
attention in statistical physics, especially in the field of spin glasses.
From a statistical physics' viewpoint, finding solutions of CSPs amounts to find the ground-state configurations of
spin glasses at zero temperature, where the energy represents the number of violated constraints.
Most interesting CSPs also display a spin glass behavior at thermodynamic limit
(with number of variables $N\to\infty$, and number of constraints $M\to\infty$), and encounters
set of phase transitions when constraint density $c=\frac{M}{N}$ increases.
The first transition that caught lots of interests is the satisfiability
transition $c_s$ \cite{Kirkpatrick-Selman-1994,Monasson-Zecchina-1999,Mezard-Parisi-Zecchina-2002} where the probability of a random instance being satisfiable changes sharply from $1$ to $0$.
In the satisfiable phase (the parameter regime
that w.h.p.\footnotemark[2] \footnotetext[2]{`with high probability'(w.h.p.) means that the probability of some event tends to 1, as $N\rightarrow \infty$.}random instances are solvable), studies using cavity method \cite{cavity,Mezard-Zecchina-2002} from spin glass theory tell us that the
solution space of CSPs are highly structured: with $c$ increasing, system undergoes
clustering transition, condensation transition and finally satisfiability transition \cite{Krzakala-etal-PNAS-2007,Montanari-Tersenghi-Semerjian-2008,color}. All of these transitions
are connected to the fact that solutions are clustered into clusters. The clustering phenomenon
is believed to effect performance of solution-finding algorithms and to be responsible for the hardness of CSPs\cite{Krzakala-etal-PNAS-2007}.

Besides heuristic analysis using cavity method, rigorous mathematical studies have also made
lots of progress on the satisfiability transitions and clustering of solutions in CSPs:
some CSP models have been proved to have satisfiability transition such as K-XORSAT and K-SAT with growing clause length; some CSP models have been proven to have a clustering phase, such as K-SAT ($K\geq 8$), K-coloring and hypergraph 2-coloring \cite{Mezard-Mora-Zecchina-2005,Achlioptas-oghlan-Tersenghi-2011,Achlioptas-oghlan-2008}. Hypergraph 2-coloring has been proven to have condensation phase in \cite{coja-oghlan-Zdeborova-2011}.

In this paper we study model RB \cite{Xu-Li-2000}, a prototype CSP model with growing domains
that is revised from the famous CSP Model B \cite{Smith-Dyer-1996}.
The main difference between model RB and classic CSPs like satisfiability problems is that
number of states (we called domain size here) one variable can take is an increasing
function of number of variables. This is probably one of the reason that makes the
satisfiability threshold rigorously solvable \cite{Xu-Li-2000}, and the clustering of
solutions also provable as we will show in the main text of this paper.
It has been shown that random instances of model RB are hard to solve
close to satisfiability transition \cite{Xu-Li-2006,Xu-etal-2007,Liu-Lin-Wang-2008,Zhao-etal-2011}, and benchmarks based on model RB (more information on \url{http://www.nlsde.buaa.edu.cn/~kexu/}) have been widely used in algorithmic research and in various kinds of algorithm competitions (e.g., CSP, SAT and MaxSAT) in recent years. Model RB has also been used or investigated in many different fields of computer science. Hardness of model RB makes the relation between its solution space structure and its hardness an interesting problem.

Using cavity method, it has been shown that \cite{Zhao-etal-2012}
replica symmetry solution is always stable in the satisfiable phase, which suggests that
condensation transition does not exist in this problem.
Here we use rigorous methods, namely the \textit{first} and the \textit{second moment method}
\cite{dimitris and moore,Achlioptas-oghlan-Tersenghi-2011},
 to show that in the
satisfiable phase close to the satisfiability transition, solutions are always
clustered into exponential number of clusters, and each cluster contains sub-exponential number
of solutions. So we are showing rigorously that the system has no condensation transition.

The main contributions of this paper are twofold:
\begin{itemize}
\item
	{From mathematical point of view, we give a rigorous analysis on the geometry of solution clusters in model RB problems.}
\item{From statistical physics point of view, we show that there is no condensation transition in this this problem. Thus as a consequence, replica symmetry results including Bethe entropy
	and marginals given by cavity method and associated Belief Propagation algorithm, should be
	asymptotically exact.}
\end{itemize}

The rest of the paper is organized as follows. Section \ref{sec:definition} includes
definitions of model RB and brief descriptions on previously obtained results
on phase transitions of model RB.
Sec. \ref{sec:solutions} contains our main results which include rigorous analysis on
clustering of solutions, number and diameter of clusters.
We conclude this work in Sec. \ref{sec:conclusion}.

\section{Model RB and phase transitions}\label{sec:definition}
Random CSP model provides a relatively ``unbiased'' samples for testing algorithms, helping design better algorithms and heuristics, provides insight into complexity theory. The standard random models (such as model B) suffer from (trivial) insolubility as problem size increases, then models with varying scales of parameters was proposed to overcome this deficiency \cite{Lecoutre-2009,smith2001,frize,fan2011,fan2012}. Model RB is one of them, who has growing domain size. It is worth mentioning that CSPs with growing domains can describe many practical problems better, for example N-queens problem, Latin square problem, sudoku, and Golomb ruler problem.

Here is the definition of model RB.
An instance of model RB contains $N$ variables, each of which takes values from its
domain $D=\{1,2,\cdots,d_N\}$, with $d_N=N^\alpha$. Note that the domain size $|D|$ is growing polynomially with
system size $N$, and this is the main difference between model RB and
classic CSPs like K-SAT problems.
There are $M=rN\ln N$ constrains in one instance, each constraint involves $k$ ($k\geq 2$)
different variables that chosen randomly and uniformly from all variables.
Total number of assignments of variables involved by a constraint is $d_N^k$.
For a constraint $a$
we pick up randomly $pd_N^k$ different assignments from totally $d_N^k$ assignments to form  an incompatible-set $Q_a$.
In other words constraint $a$ is satisfiable by the assignment
$\bre{\sigma_a}=\bre{\sigma_{a1},\sigma_{a2},...,\sigma_{ak}}$
if $\bre{\sigma_a}\not\in Q_a$.

So given parameters ($N,k,r,\alpha,p$), an instance of model RB is generated as follows
\begin{enumerate}
	\item Select (with repetition) $rN\ln N$ random constraints, each of which is formed by
		selecting (without repetition) randomly $k$ variables.
	\item For each constraint, we form an incompatible-set by uniformly select (without repetition) $pN^{\alpha k}$ elements of $D^{k}$.
\end{enumerate}
Note that here we consider
\begin{equation}
	p<1-\frac{1}{k},
\end{equation}
in order to exclude too few configurations in each constraint, and to facilitate the derivation.

Given an instance of model RB, the task is to find a solution, i.e. an assignment that
satisfies all the constraints simultaneously.
It is easy to see that total number of configurations is $N^{\alpha N}$, each of which
has probability $(1-p)^{rN\ln N}$ to satisfy all the constraints.
If we use $X$ to denote number of solutions in one instance, the expectation of
it over all possible instances can be written as
\begin{equation}
	\mathbb{E}(X)=N^{\alpha N}(1-p)^{rN\ln N}.
\end{equation}
Let
\begin{equation}
	\rcr=-\frac{\alpha}{\ln(1-p)},
\end{equation}
we can see that with $r>\rcr$, expectation of number of solutions
is nearly $0$ for large $N$.
Using Markov's inequality $$P(X>0)\leq \mathbb{E}(X),$$
we know that $\mathbb{E}(X)$ gives an upper
bound for probability of a formula being satisfiable. So for sure
with $r>\rcr$ w.h.p. there is no solution in an instance of model RB.
With $r<\rcr$, though expectation of number of solutions is larger than $0$,
these solutions may
distributed non-uniformly, that is some instances may contain exponentially many
solutions while in other instances there could be no solution at all.

Fortunately in model RB it has been shown \cite{Xu-Li-2000} that $\mathbb{E}(x)$ is square root of
expectation of the second moment of number of solutions with $r<\rcr$,
hence solutions are indeed distributed uniformly. More precisely, with $N\to\infty$,
using Cauchy's inequality, with $r<\rcr$ we have
\begin{equation}\label{eq:second_moment11}
	P(X>0)\geq \frac{E^2(X)}{E(X^2)}\rightarrow 1,
\end{equation}
In other words, the satisfiability transition happens at $\rcr$:
\begin{align}
\nonumber
\lim_{n \rightarrow \infty}{\text{Pr}(X>0)}=1\ &\text{when}\ r<\rcr\\
\nonumber
\lim_{n \rightarrow \infty}{\text{Pr}(X>0)}=0\ &\text{when}\ r>\rcr.
\end{align}
However even in the satisfiable phase close to the satisfiability transition,
where we almost sure there are
solutions, it is still difficult to find a solution in an random instance. Actually
many efforts have been devoted to designing efficient algorithms that work in this regime.
So far, our understanding on this algorithmically hardness is based on the clustering
of solutions in
the satisfiable regime close to transition. In statistical physics, the methods that we can
use to describe the solutions space structure are borrowed
from cavity method in spin glass theory.
From statistical physics point of view, CSP problems are
nothing but spin glass models at zero temperature, with energy of the system defined as
number of violated constraints in CSPs. Thus finding a solution is equivalent to finding a
configuration
$\bre{\sigma} = \bre{\sigma_i|i=1,...,N}$
that has zero energy. More precisely one can define a Gibbs measure
$$
P(\bre{\sigma})=\frac{1}{Z}e^{\beta\sum_{a=1}^ME_a(\bre{\sigma_a})},
$$
where $Z$ is partition function, $E_a(\bre{\sigma_a})$ is $0$ if ${\bre{\sigma_a}\not\in Q_a}$
and is $1$ otherwise.
By taking $\beta\to\infty$, and using e.g. cavity method,
one can study properties of this Gibbs distribution reflecting the structure of solutions space \cite{Krzakala-etal-PNAS-2007}, such as whether
ground-state energy is $0$, whether Gibbs distribution is extremal, whether replica symmetry
is broken etc.
The previous study \cite{Krzakala-etal-PNAS-2007,Montanari-Tersenghi-Semerjian-2008,color}  have shown that
the similar picture of structure of configuration space and phase transitions exist
in lots of interesting constraint satisfaction problems:
when number of constraints
is small, replica symmetry holds and Gibbs measure is extremal. With
number of constraints (or edges in the graph) increasing, system undergoes
clustering, condensation and satisfiability transitions respectively.
At the clustering
transition (also called dynamical transition), set of solutions begins to
split into exponentially number of pure states, and replica symmetry holds in each pure
state. At the condensation transition, size of clusters becomes inhomogeneous such that
a finite number of clusters contains almost all the solutions. If the number of
constraints keeps increasing and beyond satisfiability transition, neither cluster nor solution exists
any more.
Note that in some CSPs like K-SAT problem with
$K=3$ and some combinatorial optimization problems like independent set problem
\cite{vc1,vc2} with low average degree,
clustering transition and condensation transition are identical. While for some other
problems like K-SAT problem with $K\geq 4$ and graph coloring problem, condensation
transition is distinct from clustering transition, and there is a stable one step
replica symmetry breaking (1RSB) phase.

Studies based on replica symmetry cavity method and its associated Belief
Propagation equations have been applied to model RB in \cite{Zhao-etal-2012},
and Bethe entropy $S_{\text{Bethe}}$ (leading order of logarithm of number of solutions)
has been calculated on single instances.
There are two interesting observations in \cite{Zhao-etal-2012}. First, BP equations always
converge on single instances when energy reported by BP is zero. It means that
in the satisfiable phase BP is always marginally stable, indicating that replica
symmetry solution is always locally stable in satisfiable phase; Second, Bethe entropy
agrees very well with first moment estimate of entropy (annealed entropy)
$S_{\text{Bethe}}=\ln \mathbb{E}(X)$. These two phenomenons suggest that
Belief Propagation algorithm may give a asymptotically correct marginal and free energy, and condensation
transition does not exist.

\section{solution space structure of model RB}\label{sec:solutions}
Heuristic analysis on solution space structure
using cavity method and replica symmetry breaking are based on the concept of
\textit {pure state}, with assumptions of extremal Gibbs measure and exponential growth of both
number of clusters and number of solutions in each cluster.
Following \cite{Achlioptas-Ricci-Tersenghi-2006}, in this paper we use a more concrete
definition of cluster using the Hamming distance.
Let us use $\mathcal{S}$ to denote set of all solutions in an instance.
The Hamming distance between two arbitrary solutions $x,y\in \mathcal{S}$, noted $d(x,y)$,
is the number of configurations taking different values in $x,y$.
We define {\it diameter} of an set of solution $X\subseteq \mathcal{S}$ as the maximum Hamming distance between
any two elements of $X$. The distance between two sets $X,Y\subseteq \mathcal{S}$,
is the minimum Hamming distance between any $x \in X$ and any $y \in Y$.
We define {\it cluster} as a connected component of $\mathcal{S}$,
where every $x,y\in \mathcal{S}$ are considered
adjacent if they are at Hamming distance 1 (or an finite integer $q$, it does not affect the
conclusion).
We further define {\it region} as union of some non-empty clusters.

\subsection{clustering of solutions}
Our analysis is based on the number of solution pairs $Z(x)$ at Hamming distance $Nx$, with $
0<x<1$. Again as it is hard to compute $Z(x)$ exactly, we turn to the expectation of $Z(x)$.
As shown in \cite{Xu-Li-2000}, the number of assignment pairs at distance $Nx$ is equal to
$$t(x)=N^{\alpha N}\binom{N}{Nx}(N^{\alpha}-1)^{Nx};$$
probability of a pair of assignments being two solutions is written as
$$q(x)=\{(1-p)^{2}+p(1-p)[(1-x)^{k}+g(x)]\}^{rN\ln{N}},$$
where $$g(x)=\frac{-k(k-1)x(1-x)^{k-1}}{2N}.$$
So the expectation of $Z(x)$, denoted by $\mathbb{E}(Z(x))$, is the product of $t(x)$ and $q(x)$.

Since domain size grows with $N$ in model RB, it is convenient to define the normalized
version of $\mathbb{E}(Z(x))$ given $\alpha, p$ and $r$:
\begin{eqnarray}\label{eq:fx}
f(x)&=&\lim_{N\rightarrow \infty}{\ln( \mathbb{E}(Z(x)))/(N\ln N)}\nonumber\\
&=&\alpha (1+x)+r\ln {[(1-p)^2 +p(1-p)(1-x)^k]}.
\end{eqnarray}
Actually $f(x)$ is the annealed entropy density, which is a decreasing function of number
of constraints.
It is easy to see when $f(x)<0$, $\mathbb{E}(Z(x))\rightarrow 0$.

\begin{figure}
   \centering
    \includegraphics[width=0.6\columnwidth]{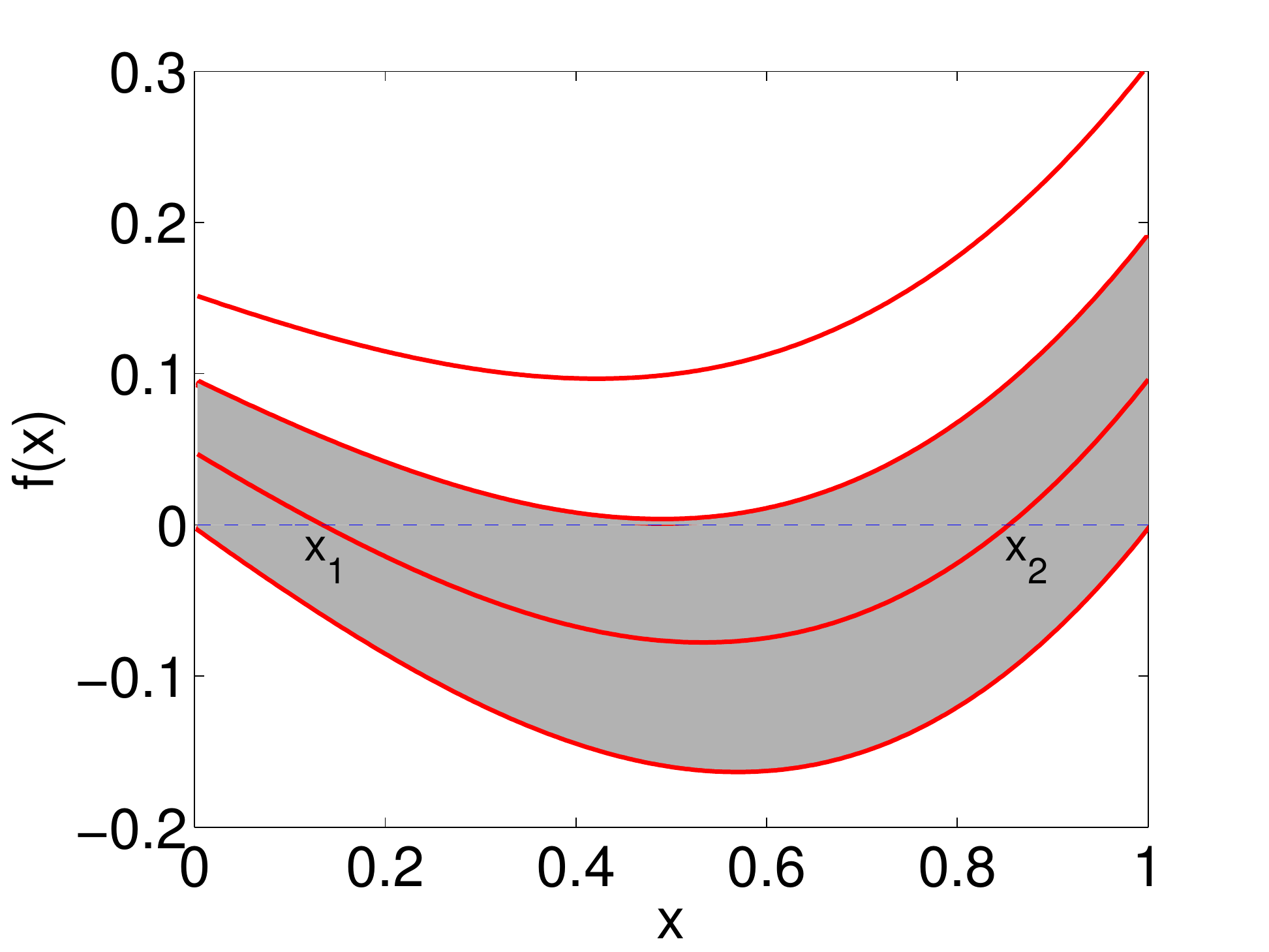}
	\caption{Annealed entropy density (leading order of logarithm of number of solution
	pairs $f(x)$ Eq.~\eqref{eq:fx}), for
	$\alpha=0.8$, $p=0.4$, $k=2$. From top to bottom, $r$ values are $0.81\rcr$,
	$0.8815\rcr$, $0.94\rcr$ and $\rcr$ respectively.
	In the shaded regime $(0.8815\rcr, \rcr)$, $f(x)=0$ has two solutions, denoted by $x_1$ and
	$x_2$. One example of the two solutions are labeled in the figure for $r=0.94\rcr$.
		\label{fig:fx}}
\end{figure}

In Fig.~\ref{fig:fx} we plot $f(x)$ as a function of $x$ for $k=2$, $p=0.4$, $\alpha=0.8$
and several different $r$ values. The top line has a relatively small $r$, we can see
that $f(x)$ is above $0$. It's worth to mention this does not mean there are exponential
number of solutions at distance $Nx$, because $f(x)$ is only a lower bound
for $\ln( Z(x))/(N\ln N)$, indicated by Markov's inequality
$$P(Z(x)\geq 0)\leq\mathbb E(Z(x)).$$
With $r$ increasing, this $f(x)$ curve becomes lower and lower.
At a certain value $\hat r$, $=0.8815\rcr$ in our example in Fig.~\ref{fig:fx},
$f(x)$ curve reaches $0$. Beyond $\hat r$, $f(x)=0$ has $2$
solutions \footnotemark[3]\footnotetext[3]{There are at most $2$ solutions, following
the concavity of $f(x)$ shown in Appendix.} until $r$ reaches $\rcr$.
With $r>\rcr$, it has been proved \cite{Xu-Li-2000} that there is no solution in the system
which is consistent with what the curve shows: the upper bound of $Z(x)$ becomes negative
for any $x$ value.

We focus on the regime between $\hat r $ and $\rcr$ (shaded regime in the figure) when
$f(x)=0$ has two solutions, denoted by $x_1$ and $x_2$.
Using definition of $f(x)$ in
Eq.~\eqref{eq:fx}, we can compute number of solutions at Hamming distance between
$x_1$ and $x_2$ with $N\to\infty$:
\begin{eqnarray}
	P\bra{\sum_{xN=x_1N+1}^{x_2N-1}Z(x)\geq 0}&\leq& \mathbb{E}\left(\sum_{xN=x_1N+1}^{x_2N-1}{Z(x)}\right)\nonumber\\
&\leq& (x_2-x_1)N\cdot \max_{x\in (x_1,x_2)}{\mathbb{E}\left(Z(x)\right)}\nonumber\\
&\rightarrow &(x_2-x_1)N\cdot \max_{x\in (x_1,x_2)}{N^{f(x)N}}\nonumber\\
&\rightarrow& 0.
\end{eqnarray}
Thus the last equation indicates that w.h.p. there is no solution pair at Hamming distance between
$x_1$ and $x_2$.

On the other hand, with $N\to\infty$, using Paley-Zigmund inequality we have
\begin{eqnarray}\label{eq:second_moment}
P[ X>\frac{1}{N}\mathbb{E}(X) ] &\geq& \frac{\left(\mathbb{E}(X)-\frac{1}{N}\mathbb{E}(X)\right)^2}{\mathbb{E}(X^2)}\nonumber\\
&\rightarrow&(1-\frac{1}{N})^2\rightarrow 1,
\end{eqnarray}
where we have made use of Eq.~\eqref{eq:second_moment11} that
$\lim_{N\to\infty}\frac{(\mathbb{E}(X))^2}{\mathbb{E}(X^2)}= 1$. The number of solutions $X$ is w.h.p. bigger than its mean divided by $N$.

It follows that w.h.p.
in the regime where ($x_1$, $x_2$) pair exists (e.g. shaded regime in Fig.~\ref{fig:fx}), system has exponentially number of solutions and their Hamming
distance is discontinuously distributed. In other words, the solution space is clustered.
Actually we can show that for all parameters of model RB, there always exists such
clustered regime.
A proof for the existence of $x_1$ and $x_2$ pair is given in Appendix \ref{sec:concavity}.

\subsection{Organization of clusters}
In this section we give a precise description on clustering of solutions, including
bounds for diameter of clusters, distance between clusters, number of solutions in one cluster and
number of clusters in the satisfiable phase.
Given the result from last section, using method from Achlioptas and Ricci-Tersenghi
(referring to section 3 of \cite{Achlioptas-Ricci-Tersenghi-2006}, \cite{Achlioptas-oghlan-Tersenghi-2011}, section 3 of \cite{Achlioptas-2008}), we actually have a concrete way to split
the solution space and put solutions into different clusters:
Assuming we know all the solutions, we can split the solution space by the
cured surface $\{y|d(x,y)= x_1 N\}$ for each solution, and obtain
a set of regions $\mathbb A$. In more detail we can do it as follows:
\begin{enumerate}
\item  Initialize $\mathbb A =\{\mathcal{S}\}$, with $\mathcal{S}$ denoting the set of all solutions.
\item  For every solution $x\in \mathcal{S}$, repeat splitting step (step 3) around $x$.
\item  Splitting step: denote the only region including $x$ in $\mathbb A$ by $A$. If there is $y\in A$ satisfied $d(x,y)> x_1 N$, then let $B=\{y|y\in A,d(x,y)\leq x_1 N\}$, $C=A\setminus B$, and let $\mathbb A=(\mathbb A \cup \{B\} \cup \{C\})\setminus \{A\}$.
\end{enumerate}
The final $\mathbb A$ is the set of regions we want. We can show that
$\mathbb A$ has following properties:
\begin{itemize}
\item
	{The diameter of each region is at most $x_1 N$. Because if there are two solutions at distance more than $x_1 N$ in a region, splitting step will for sure split them
	into different regions.}
\item{
	The distance between every pair of regions is at least  $(x_2 - x_1 )N$.
	To show this, assume there are three solutions $x$, $y$ and $z$, they are put into two clusters
	after the splitting step around $x$: $y$ is put to the same region with $x$, and
	$z$ is put to a different region. Then we have $d(x,y)\leq x_1N$, $d(x,z)\geq x_2N$,
	and the triangle rule implies that $d(y,z)\geq (x_2 - x_1)N$. }
\end{itemize}

Another important property we are interested in is the number of solutions in clusters. Here for convenience we talk about typical instances of model RB, only to avoid repeatedly using of ``w.h.p.''.
From above analysis we know that the diameter of each region is at most $x_1N$, so
number of solution pairs in one cluster is bounded above by number of solution pairs
Hamming distance smaller than $x_1$. Letting
\begin{equation}
	l=\max_{x \in [\frac{1}{N},x_1]}{\mathbb{E}(Z(x))}
\end{equation}
and using Markov's inequality
we have
\begin{eqnarray}
	P\left( \sum_{xN =1}^{x_1N}{Z(x)} \geq N^2l \right) \leq \frac{\sum_{xN = 1}^{x_1N}{\mathbb{E}(Z(x))}}{N^2l} \leq \frac{N l}{N^2 l}=\frac{1}{N}.
 \end{eqnarray}
We can see that every region in $\mathbb A$ have at most $N^2l$
pairs of solutions, which implies that every region in $\mathbb A$ have at most
$N\sqrt{l}$ solutions.
We know that $f(x)$ is a concave function, and is
monotonically decreasing with $x\in[0,x_1]$ (see Appendix~\ref{sec:concavity} for a proof),
so as $N$ is very large, number of solutions in one cluster is smaller than
$$N\sqrt{l}< N^{1/2[\alpha+r\ln(1-p)]N+1}.$$

Note that compared with the lower bound of total number of solutions
$\frac{1}{N}\mathbb{E}(X)$ (Eq.\eqref{eq:second_moment}),
number of solutions in one region is exponentially smaller.
To make it more precise, if we define a complexity function $\Sigma$
representing leading order of logarithm of number of clusters divided by $N\log N$
(note that in our system, correct scaling for densities is $N\log N$), we have as $N$ is very large
\begin{eqnarray}\label{eq:complexity}
	\Sigma &\geq&
	\frac{1}{N\ln N}\ln \bra{\frac{ \frac{1}{N}\mathbb{E}(X)}{N\sqrt{l}}}\nonumber\\
	&\geq& \frac{1}{2}\brb{\alpha+r\ln(1-p)}-\frac{2}{N}.
\end{eqnarray}
Last equation says that in the satisfiable phase, complexity is positive all the way down to the transition.

A direct implication from above results is that in whole parameter range,
phase diagram of model RB does not contain condensed clustered phase, because there does not
exist a set of finite number of clusters that contain almost all the solutions.
In replica symmetry breaking theory, existence of clustering transition is indicated by
$\hat \Sigma(m=1)>0$ and existence of condensation is indicated by
$\hat \Sigma(m=1)<0$ where $\hat\Sigma$ denotes complexity
which is leading order of logarithm of number of pure states as a function of Parisi
parameter $m$ \cite{Krzakala-etal-PNAS-2007}. With Parisi parameter $m=1$, first step replica symmetry
breaking solution actually gives equal weight to each pure state, thus the total free energy
is identical to the replica symmetry free energy. We can see that our definition of
complexity $\Sigma$ is very similar to $\hat \Sigma(m=1)$ because it gives equal weight to
different clusters. Thus $\Sigma>0$ all the way down to the satisfiability transition is
another way to show that there is no condensation transition in model RB.
Note since our definition of clusters is different from pure state (as we do not refer to
properties of Gibbs measure), our claim is not a proof.

\section{conclusion and discussion}\label{sec:conclusion}
As a conclusion, in this paper we described in detail the solution space structure of
model RB problem using rigorous methods. We show that close to the satisfiability transition,
solutions clustered into exponential number of clusters, each of which contains sub-exponential
number of solutions. And we showed that there is no condensation transition in model RB which
testifies an statement of Zhao et al \cite{Zhao-etal-2012} using non-rigorous
cavity methods from statistical physics.

The factor graph of model RB has a special feature that the link degree per variable is very large (growing with number of variables $N$), which is the same as model K-SAT with growing K \cite{Fri}. We think this feature will affect phase transitions, and we will put more thoughts on that in future work.

We note that though we proved the clustering of solutions close to the satisfiability transition, we
are still not sure where the clustered phase begins. Though lack of rigorous methods,
heuristically the clustering transition can be
estimated when one step replica symmetry breaking cavity method at Parisi
parameter $m=1$ begins to have non-trivial solution.
We will address this point in future work.

It has been shown that instead of clustering, freezing of
clusters is the real reason for algorithmic hardness. Numerical experiments made in
\cite{Zhao-etal-2012} and \cite{Zhao-etal-2011} showed that starting from $\hat r$ (where $f(x)=0$ has only one
solution), the most efficient
algorithms begin to fail in finding solutions, so it suggests that
clusters become frozen immediately at $\hat r$. This would be interesting to
study in detail.

\section{Acknowledgments}
Pan Zhang wishes to thank Cristopher Moore for helpful conversations.

\begin{appendix}
\section{Concavity of $f(x)$}\label{sec:concavity}
The first and second derivatives of $f(x)$ with respect to $x$ read
\begin{eqnarray}
	\frac{\partial f}{\partial x}&=&\alpha-\frac{ rpk(1-x)^{k-1}}{ 1-p+p(1-x)^k}\nonumber\\
\frac{\partial^2 f}{\partial x^2} &=& \frac{  rpk}{\brb{ 1-p+p(1-x)^k}^2}
\brb{ (k-1)(1-p)(1-x)^{k-2}-p(1-x)^{2k-2}}.\nonumber
\end{eqnarray}
Then it is easy to check that $\frac{\partial^2 f}{\partial x^2}$ is always positive for $x\in[0,1]$
with $k\geq 2$ and $p\leq 1-\frac{1}{k}$, which implies the concavity of $f(x)$.

Observe that both $f(0)$ and $f(1)$ are positive in the satisfiable phase,
$f(0)=f(1)=0$ at the satisfiable-unsatisfiable transition, and
$\left . \frac{\partial f(x)}{\partial x}\right |_{x=1}=\alpha >0.$
So using the concavity of $f(x)$, it is obvious that there must
exist $r<\rcr$, and pair $x_1, x_2$ such that $f(x)<0$ with $x_1<x<x_2$.
Moreover in the satisfiable phase, given
$f(0)=\alpha+r\ln(1-p)>0$ and $x_1$ is the first point that $f(x)$ reaches $0$,
we can conclude that $f(x)$ is a
monotonically decreasing function with $x\in [0,x_1]$.

\end{appendix}


\end{document}